\documentclass[useAMS,usenatbib]{mnras}
\usepackage{newtxtext,newtxmath}
\usepackage[T1]{fontenc}
\usepackage{ae,aecompl}
\usepackage{graphicx}	           
\usepackage{amsmath}	
\usepackage{amssymb}	
\usepackage{natbib}
\usepackage[para,online,flushleft]{threeparttable}
\usepackage{booktabs}
\usepackage{color}
\definecolor{modif}{rgb}{0,0.7,0}
\definecolor{comment}{rgb}{1,0,0}
\newcommand{\kms}{km\,s$^{-1}$}

\newcommand{\GG}[1]{}

\definecolor{darkblue}{rgb}{0.0, 0.0, 0.60}
\usepackage{hyperref}
\hypersetup{bookmarks=true, colorlinks=true, linkcolor=darkblue, citecolor=darkblue, filecolor=darkblue, urlcolor=darkblue}

\newcommand{\bz}{\ensuremath{\langle B_z\rangle}}
\newcommand{\nz}{\ensuremath{\langle N_z\rangle}}

\newcommand{\HD}{HD\,54879}

\def\gtrsim{\mathrel{\hbox{\rlap{\hbox{\lower4pt\hbox{$\sim$}}}\hbox{$>$}}}}
\def\ltsim{\mathrel{\hbox{\rlap{\hbox{\lower4pt\hbox{$\sim$}}}\hbox{$<$}}}}

\title[No sudden change of \HD]{No evidence of a sudden change of spectral appearance or magnetic field strength of the O9.7V star \HD}

\author[G.A. Wade et al.]{G.A. Wade\thanks{E-mail: wade-g@rmc.ca}$^1$, S. Bagnulo$^2$, Z. Keszthelyi$^{3}$, C.P. Folsom$^{4}$, E. Alecian$^{5}$,  
\newauthor{N. Castro$^{6}$, A. David-Uraz$^{7}$, L. Fossati$^{8}$, V. Petit$^7$, M.E. Shultz$^7$, J. Sikora$^{9}$}
\\
$^{1}$Dept. of Physics \& Space Science, Royal Military College of Canada, PO Box 17000 Station Forces, Kingston, ON, Canada K7K 0C6 \\
$^2$Armagh Observatory and Planetarium, College Hill, Armagh BT61 9DG, UK  \\
$^3$Anton Pannenkoek Institute for Astronomy, Universiteit van Amsterdam, Science Park 904, NL-1098 XH Amsterdam, the Netherlands\\
$^{4}$IRAP, Universit\'{e} de Toulouse, CNRS, UPS, CNES, 14 Avenue Edouard Belin, 31400, Toulouse, France\\ 
$^{5}$Universit\'e Grenoble Alpes, IPAG, CNRS, F-38000 Grenoble, France \\
$^{6}$Leibniz-Institut f\"r Astrophysik Potsdam, An der Sternwarte 16, 14482 Potsdam, Germany \\ 
$^7$Dept. of Physics and Astronomy, University of Delaware, 217 Sharp Lab, Newark, DE 19716, USA\\
$^{8}$Space Research Institute, Austrian Academy of Sciences, Schmiedlstrasse 6, A-8042 Graz, Austria \\ 
$^9$Dept. of Physics and Astronomy, Bishop's University, Sherbrooke, Qu{\'e}bec, Canada, J1M 1Z7
}

\begin{document}

\date{Accepted . Received , in original form }

\pagerange{\pageref{firstpage}--\pageref{lastpage}} \pubyear{2002}

\maketitle

\label{firstpage}

\begin{abstract}
It was recently claimed that the magnetic O-type star \HD\ exhibits important radial velocity variability indicative of its presence in a spectroscopic binary. More remarkably, it was furthermore reported that the star underwent a short, sudden variation in spectral type and magnetic field. In this Letter we examine new Narval and ESPaDOnS data of this star in addition to the previously-published FORS2 data and conclude that both the reported velocity variations and the sudden spectral and magnetic changes are spurious. 
\end{abstract}


\section{Introduction}

\HD\ is a late O-type star that was reported to be magnetic by \citet{castro2015}. They reported that the spectrum, characterised by very sharp lines, was very stable and essentially unperturbed by the presence of a magnetosphere. \citet{shenar2017} reported multiwavelength (optical, UV, X-ray) observations of \HD, deriving an effective temperature $T_{\rm eff} = 30.5 \, \pm 0.5 \, \mathrm{kK}$ and a surface gravity of $\log g = 4.0 \pm 0.1$. Both  \citet{castro2015} and \cite{shenar2017} concluded that the magnetic and spectral variability imply a rather long rotation period, likely around 5 years. Hence \HD\ is inferred to be one of the most slowly-rotating O-type stars known.

\citet{hubrig2019} described their analysis of a FORS2 spectropolarimetric timeseries of this star spanning 140 days. The essential conclusions of their study can be summarized as follows: (i) HD\,54879 exhibits significant (of order 100\,\kms) radial velocity variations on timescales of days. (ii) On JD 2458166, HD\,54879 underwent ``a sudden, short-term increase of the magnetic field strength" (from about -100\,G to -800\,G), accompanied by a remarkable change in the star's spectrum corresponding to a significant change in spectral type (from late O to early B, accompanied by the complete disappearance of the He\,{\sc ii} lines).  In a very recent follow-up paper and erratum, \citet{hubrig2019b,hubrig2019c} appear to link both the sudden change of the spectral appearance of \HD\ and the radial velocity variation to an insufficient S/N of the FORS spectra, and refer to putative instabilities of the pipelines they have attempted to use, but ultimately fail to give a satisfactory explanation on how ``an imperfect spectral extraction" could transform a O9.7V-type spectrum into a B2V-type spectrum.

In this Letter we examine these claims in the light of independent spectropolarimetric observations obtained using the high-resolution spectropolarimeters Narval and ESPaDOnS, as well as a re-examination of the FORS2 data studied by \citet{hubrig2019}. 

\section{Observational material}

\begin{figure}
\includegraphics[width=8.5cm]{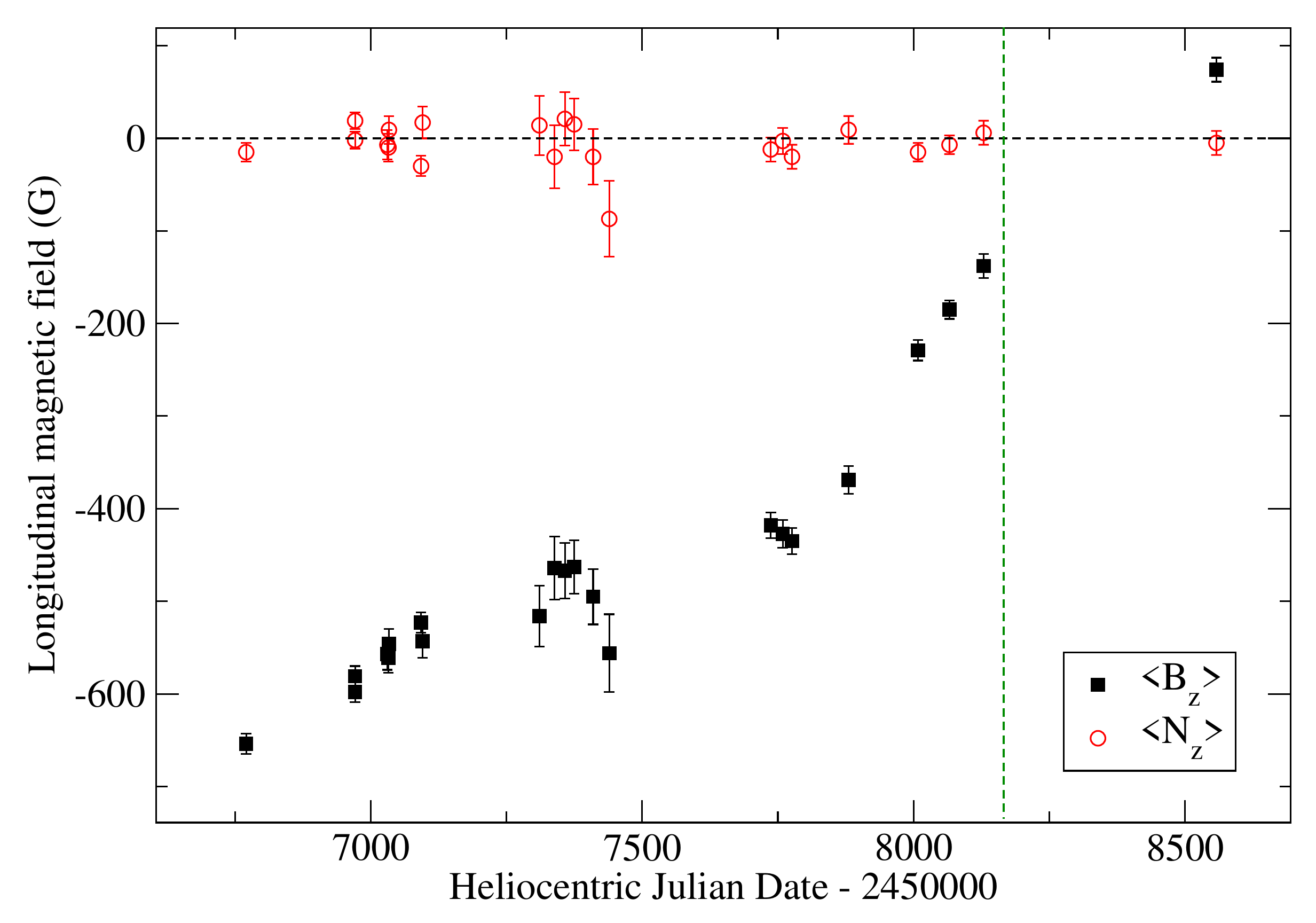}\\
 \caption{Longitudinal magnetic field of \HD\ observed with the ESPaDOnS, Narval, and HARPSpol high resolution spectropolarimeters during the period April 2014-March 2019 (4.9 years). The filled black squares represent the longitudinal field measured from the Stokes $V$ profile. The open red circles represent the longitudinal field measured from the diagnostic null ($N$) profiles. The horizontal black line indicates $\bz=0$. The vertical green line indicates JD 2458165, the date on which \HD\ was claimed to have undergone a qualitative change of its spectrum and magnetic field.\label{bz}} 
 \end{figure}

We obtained 22 high resolution circularly polarized (Stokes $V$) spectra between November 2014 and January 2018 in order to confirm the detection of the star's magnetic field, and to measure the longitudinal field variation. The observations of \HD\ were obtained using the ESPaDOnS and Narval spectropolarimeters located at the Canada-France-Hawaii Telescope (CFHT) and the Bernard Lyot Telescope (TBL), respectively. ESPaDOnS and Narval are essentially identical instruments consisting of high resolution (resolving power R$\sim$65000) \'echelle spectrographs, which are fibre-fed from Cassegrain-mounted polarimetric modules.
Each Stokes $V$ observation consisted of a sequence of 4 sub-exposures, between which the polarimetric optics of the instruments were rotated, allowing for the removal of instrumental systematics (see e.g. \citealt{donati1997}). Exposure times were adjusted for each observing run at either telescope, with sub-exposure times ranging between 475 and 900 s. The peak signal-to-noise ratio (S/N) per spectral pixel ranged from about 165 to 825. The log of observations is reported in Table~\ref{data}.
The data were reduced using pipelines specific to the CFHT and TBL, both feeding the same underlying reduction code, Libre-ESpRiT \citep{donati1997}.
We combined the new data with the 3 archival Stokes~$V$ spectra of HD\,54879 obtained by \cite{castro2015} using the HARPSpol instrument (R$\sim$115000) of the European Southern Observatory (ESO) 3.6 m telescope at La Silla Observatory. 

Each spectrum was processed using Least-Squares Deconvolution \citep[LSD;][]{donati1997} using the iLSD approach of \citet{2010A&A...524A...5K}. The line mask was developed using an Vienna Atomic Line Database \citep[VALD; e.g. ][]{1995A&AS..112..525P} {\sc extract stellar} request that was then 'cleaned' and 'tweaked' to best match the spectrum of \HD\ \citep[e.g.][]{grunhut2017}.

In addition, we have downloaded from the ESO archive the FORS2 observations discussed by \citet{hubrig2019}. Data were reduced using dedicated IRAF and FORTRAN routines as explained for instance by \citet{2018A&A...618A.113B}. We remind the reader that FORS2 is a Cassegrain-mounted instrument, and that FORS calibrations are obtained the day after the observations, with the telescope pointing at zenith. Because of unavoidable, variable flexures, radial velocity measurements require that FORS spectra be corrected for the resultant shifts using telluric lines, in particular sky emission lines. The FORS observations were obtained with grism 600B ($R\sim2000$), and the observed spectral range includes only one appropriate calibration line, the O\,{\sc i} line at 5577\,\AA. Therefore, only an approximate absolute wavelength calibration is possible. Nevertheless, using the ESO FORS pipeline \citep{2010SPIE.7737E..29I}, we obtained 2-D wavelength-calibrated frames, and measured the position of the telluric O\,{\sc i}\, 5577\,\AA\ line on each individual frame to roughly correct (to a precision of order 10\,\kms) for flexures and other instrument-related systematics. 


\begin{table}
\centering
\begin{tabular}{cccrrcrrr}   
\hline
HJD & $t_{\rm exp}$  & SNR & $B_{\rm z}$ & $N_{\rm z}$ & Inst. \\ 
-2450000 & [s]  & [/pix] &  [G] &  [G] & \\ 
\hline 
%
6770.4993   & 2700 & 384   &  $ -654  \pm    11 $  &  $  -15   \pm     10 $    & H \\
6971.0658   & 840  & 822   &  $ -581  \pm    11 $  &  $   -2   \pm      9 $    & {E}  \\
6971.1066   & 840  & 823   &  $ -598  \pm    11 $  &  $   19   \pm      9 $     & {E}  \\
7030.5195  & 900  & 391   &  $ -557  \pm    17 $  &  $   -7   \pm     16 $  & {N}  \\
7032.5133  & 900  & 411   &  $ -561  \pm    16 $  &  $  -10   \pm     15 $  & {N}  \\
7033.5159  & 900  & 413   &  $ -546  \pm    16 $  &  $    9   \pm     15 $  & {N}  \\
7092.5485   & 1800 & 212   &  $ -523  \pm    11 $  &  $  -30   \pm     11 $    & H \\
7095.5057   & 900  & 130   &  $ -543  \pm    18 $  &  $   17   \pm     17 $   & H \\
7310.6113  & 590  & 211   &  $ -516  \pm    33 $  &  $   14   \pm     32 $ & {N}     \\
7338.6680  & 590  & 239   &  $ -464  \pm    34 $  &  $  -20   \pm     34 $ & {N}     \\
7357.6202  & 590  & 200   &  $ -467  \pm    30 $  &  $   21   \pm     29 $ & {N}     \\
7374.5860  & 590  & 224   &  $ -463  \pm    29 $  &  $   15   \pm     28 $ & {N}     \\
7409.5524  & 590  & 227   &  $ -495  \pm    30 $  &  $  -20   \pm     30 $   & {N}   \\
7439.4167  & 590  & 165   &  $ -556  \pm    42 $  &  $  -87   \pm     41 $ & {N}     \\
7736.9843  & 475  & 556   &  $ -418  \pm    14 $  &  $  -12   \pm     13 $   & {E}   \\
7758.8779   & 656  & 542   &  $ -427  \pm    15 $  &  $   -3   \pm     14 $   & {E}   \\
7775.9878   & 840  & 606   &  $ -435  \pm    14 $  &  $  -20   \pm     13 $     & {E}  \\
7880.2244   & 712  & 558   &  $ -369  \pm    15 $  &  $    9   \pm     15 $   & {E}   \\
8008.1272  & 880  & 626   &  $ -229  \pm    11 $  &  $  -15   \pm     10 $   & {E}   \\
8066.0351  & 880  & 526   &  $ -185  \pm    10 $  &  $   -7   \pm     10 $   & {E}   \\
8128.9253  & 880  & 406    &  $ -138  \pm    13 $  &  $    6   \pm     13 $ & {E}  \\
8557.8753  & 880  & 506    &  $   74  \pm    13 $  &  $   -5   \pm     13 $ & {E}  \\\hline\hline
\end{tabular}
 \caption{ \label{data}Log of high-resolution spectropolarimetric observations of \HD, including previously-published HARPSpol observations (E=ESPaDOnS, N=Narval, H=HARPSpol). An integration range of +4 to +50 km/s was used to measure the longitudinal magnetic field.}
%
\end{table}

\section{Results from the ESPaDOnS and Narval observations}

\begin{figure}
 \includegraphics[width=8.5cm]{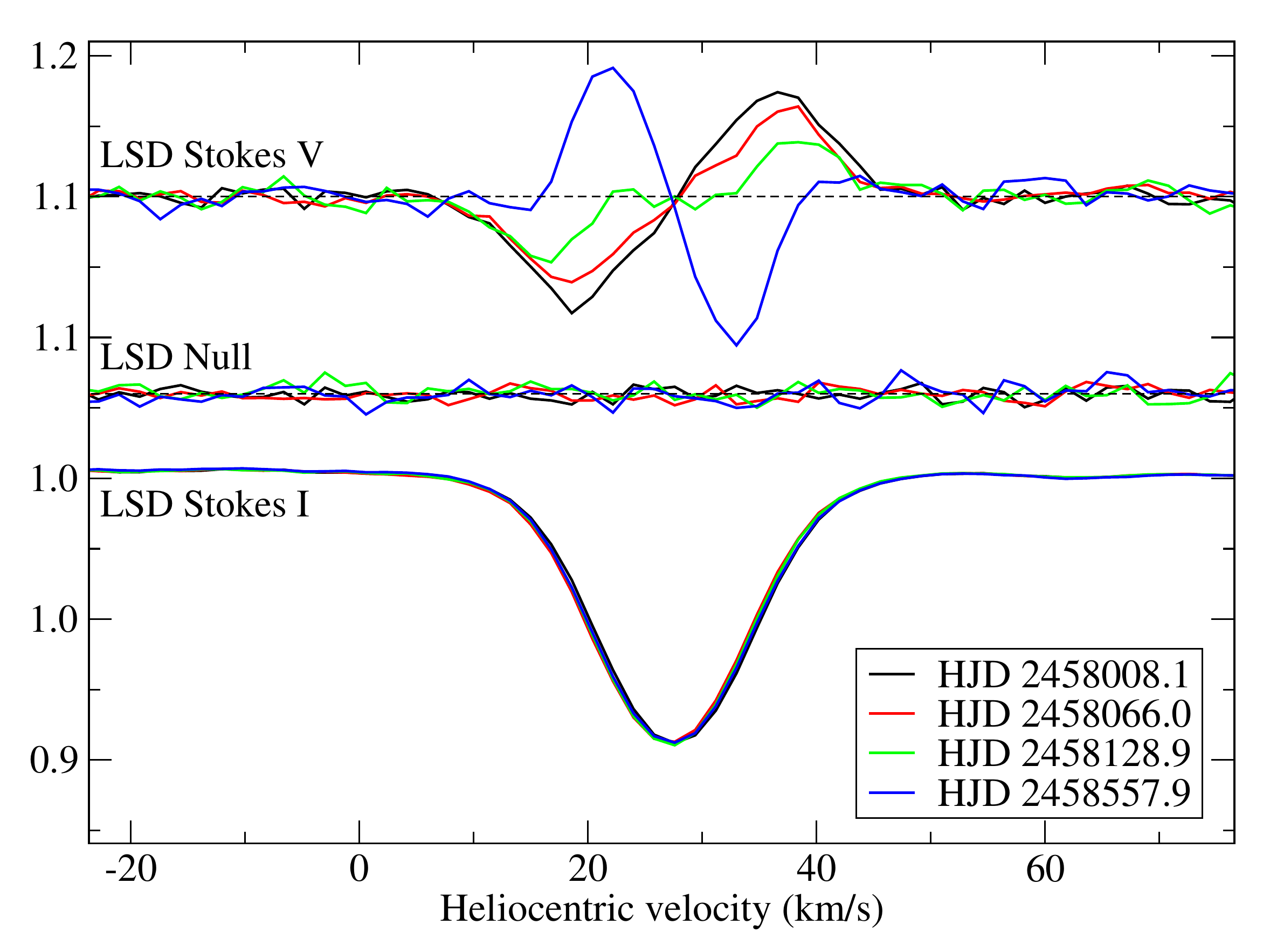}\\
 \caption{Stokes $I$ and $V$, along with null $N$ LSD profiles of \HD\ obtained in 2017-2019, illustrating the recent change of polarimetry of the Stokes $V$ signature. These spectra correspond to the final 4 longitudinal field measurements shown in Fig.~\ref{bz}.\label{stokes}} 
 \end{figure}

The observations summarized in Table~\ref{data} span nearly 5 years, including the dates during which the FORS2 observations were obtained by \citet{hubrig2019}. The derived values of the longitudinal magnetic field (\bz) are reported in this table, and were measured using the procedure described by e.g. \citet{wade2000}. As illustrated in Fig.~\ref{bz}, \bz\ (shown as black, filled points) increases more-or-less monotonically over the 5-year span of the observations, while the equivalent measurements from the diagnostic null profiles (\nz) remain consistent with zero. From this figure we draw the following conclusions:

\noindent (i) The slow, monotonic decrease of the star's longitudinal magnetic field \citep[first reported by][]{hubrig2019,hubrig2019b} shows clearly that the rotational period of \HD\ must be significantly longer than 5\,years \citep[according to the Oblique Rotator Model; ][]{1950MNRAS.110..395S}. In fact, the most recent observation obtained with ESPaDOnS (in May 2019) has (finally!) shown the reversal of sign of the Stokes $V$ profile corresponding to our first view of the northern magnetic hemisphere of the star (Fig.~\ref{stokes}). 

\noindent (ii) The high-resolution magnetic measurements show no evidence of any change of character either before or after the reported sudden change in magnetic field strength (the date of which is indicated by the vertical dashed green line in Fig.~\ref{bz}). The high-resolution measurements bracket this date, and show no departure from the long-term trend of increasing longitudinal field strength.

\begin{figure}
 \includegraphics[width=8.5cm]{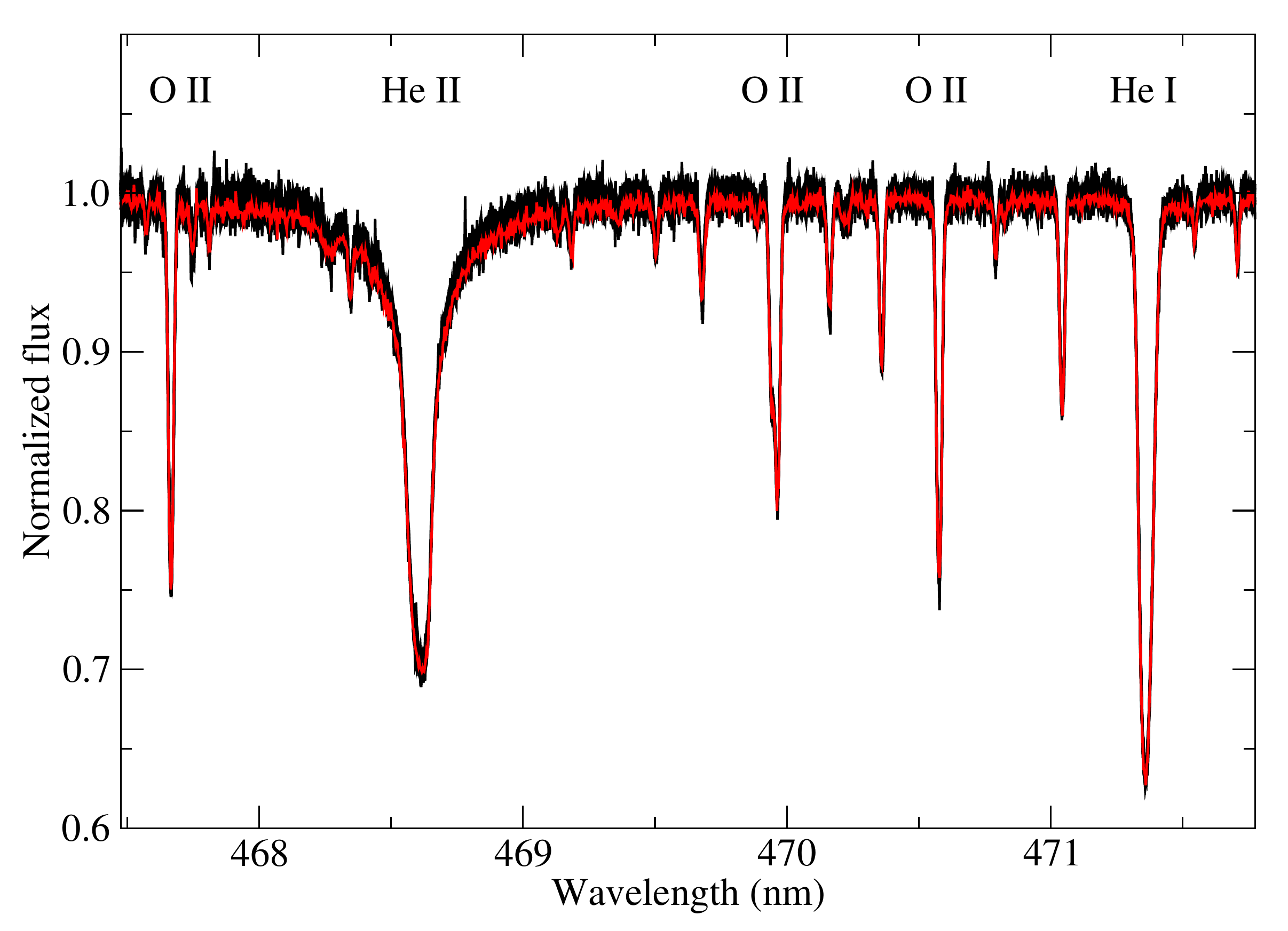}\\
 \caption{Selected ESPaDOnS and Narval spectra of \HD\ obtained between November 2014 and March 2019 (in black), showing an arbitrary region containing various absorption lines. No strong RV variability comparable to that reported by \citet{hubrig2019} is observed. The spectrum shown in red corresponds to JD 2458128.9, and was acquired about one month before the remarkable spectrum reported by \citet{hubrig2019}.\label{spec}} 
 \end{figure}

\noindent(iii) An ESPaDOnS observation of \HD\ was obtained on JD 2458129, approximately one month prior to the reported episode. As illustrated in Fig.~\ref{spec}, the spectrum obtained on that date is fully compatible with the high-resolution spectrum observed at all earlier dates, modulo very weak variations of the line profile depth and morphology (variations that are observed throughout the 5 years of monitoring). In particular we note the lack of any measurable variation of the radial velocity, in the ESPaDOnS and other high-resolution spectra (some of which were acquired less than one night apart), larger than about 1~km/s. In addition, the following ESPaDOnS spectrum (obtained on HD 2458557) is also in good agreement with the earlier data. These results are fully consistent with the initial report by \citet{castro2015}.

\section{Re-analysis of the FORS2 spectra}

\begin{figure*}
 \includegraphics[width=14cm,angle=-90]{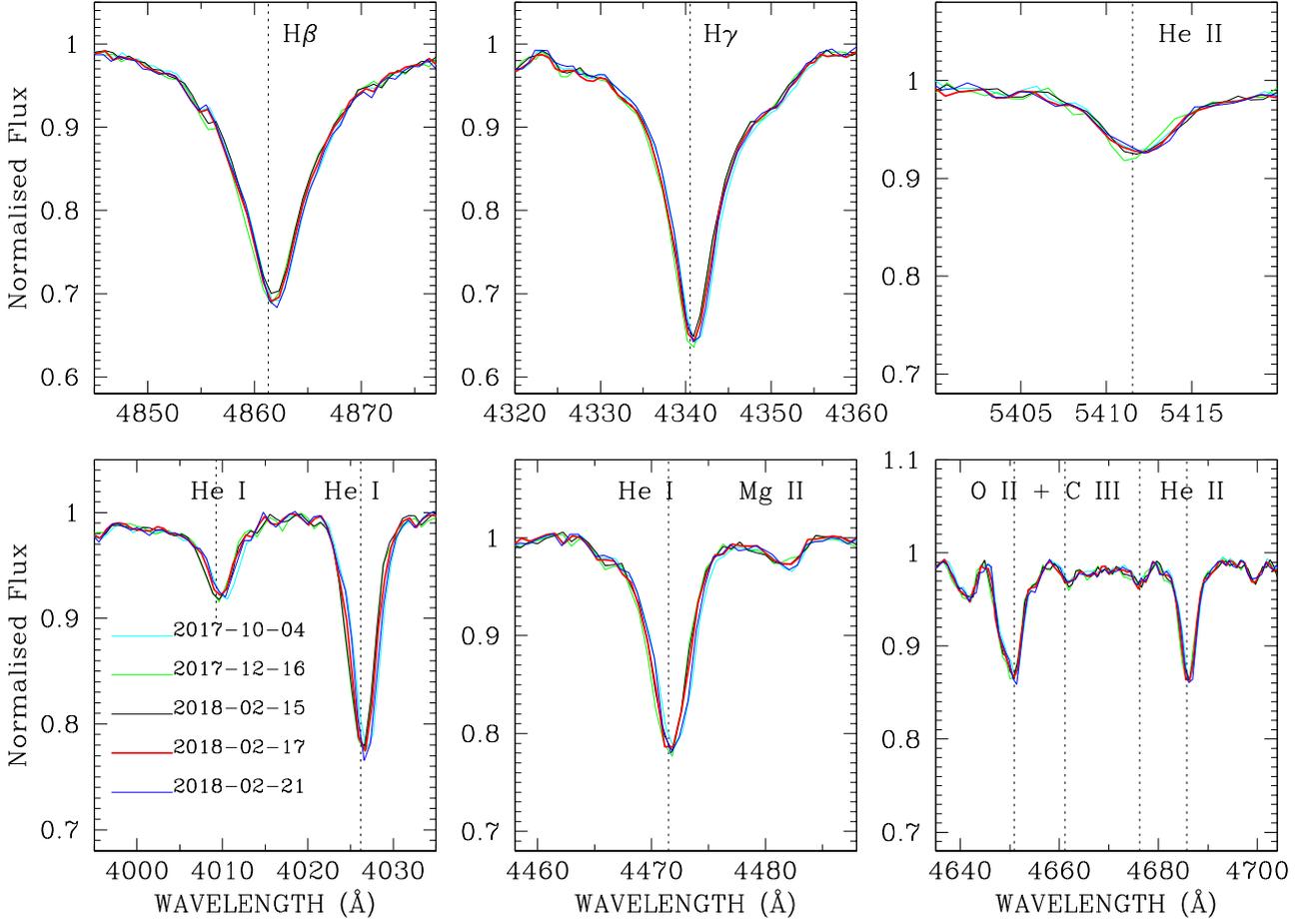}\\
 \caption{Profiles of various spectral lines in the FORS2 spectra of HD~54879 after heliocentric correction and correction for instrumental systematics using the O\,{\sc i}\,5577\,\AA\ line. }\label{forsspec}
 \end{figure*}

The major result of our re-analysis of the FORS2 spectra is that, in contrast to the claims of \citet{hubrig2019}, there is no evidence of any significant change in the spectrum obtained on February 17, 2018 with respect to all the other spectra of the same star, and that, consistent with what we have found with ESPaDOnS observations, there is no indication of any remarkable change of the star?s radial velocity. In Fig.~\ref{forsspec} we have tried to reproduce the dramatic spectral difference
shown in figure 4 of \citet{hubrig2019}. Our figure clearly shows that, not only there is no remarkable difference between the line profiles obtained at different epochs (including those obtained on February 17, 2018, shown in bold red), but that there is no significant change in radial velocity, once corrections due to the heliocentric velocity and instrument flexures are properly taken into account. We have also re-calculated the longitudinal magnetic field from all FORS2 observations using the methodology of e.g. \citet{2012A&A...538A.129B}. We found that our uncertainties are often two times larger than those published by \citet{hubrig2019}, but the field measurements appear consistent within the error bars. The only exception, of course, is the field estimate obtained from the observations of February 17, 2018, for which we have measured a longitudinal field value of $-250 \pm 140$\,G, instead of $-880 \pm 120$\,G as reported by \citet{hubrig2019}. 

\citet{hubrig2019b} and \citet{hubrig2019c} also seem to come to conclusions similar to our own: namely that the reports of the sudden spectral and magnetic change and the large velocity variations are erroneous. However, they state that low S/N of the data is the culprit, and claim that this phenomenon is a reproducible, albeit spurious consequence of a problem with the reduction pipelines used.Their figure~8 shows a $\sim 300$~\AA\ region of two FORS2 observations obtained on 1 January, 2019 separated by just 25 minutes. One of the spectra, extracted using a proprietary MIDAS pipeline, completely lacks various absorption lines that are clearly visible in the other spectrum. While the data obtained in January 2019 are not publicly available, it seems unlikely that low S/N affects spectral extraction in a way that absorption lines are filled to the continuum; moreover, a Stokes $I$ spectrum with a S/N of about 1000 cannot be considered to be ``low". In their figure~10, \citet{hubrig2019b} show some details of Stokes $I$ spectra extracted using the ESO pipeline, in which the He\,{\sc i}\, $4921$~\AA\ line appears at different wavelengths, offset by several angstroms, during a full sequence of sub-exposures obtained on a timescale of tens of seconds. Their stated conclusion is that ``the ESO FORS pipeline has issues with wavelength stability even for higher S/N data" and that ``in its current form [it] is not delivering proper results". We do not have access to the raw data presented in their figures~8 and 10, nor do the authors provide details about how they have set up or employed the pipeline. All we can say is that, to the best of our knowledge and based on extensive experience, it is unlikely that the ESO pipeline changes the applied wavelength solution between observations obtained sequentially, unless specifically set to do so by the user. In fact, using the spectra that are already available in the ESO archive, we have verified that the ESO FORS pipeline does not produce the spurious results presented by \citet{hubrig2019b}. We also note that each of the spectra obtained on 17 February, 2018 have typically a peak ADU count of 20,000 per CCD pixel, and that the combined spectrum reaches a S/N of 1600 per \AA. It is totally unclear how this spectrum could be defined having a ``low S/N", especially considering that the spectrum must have been repeatedly and carefully re-examined by \citet{hubrig2019b} and \citet{hubrig2019c}.

 
\section{Conclusions}

Motivated by the report of remarkable magnetic and radial velocity behaviour of the magnetic O-type star \HD\ by \citet{hubrig2019}, we have examined magnetic and spectral measurements of this object obtained with multiple instruments over a 5-year period. While we confirm the slow increase of the longitudinal field reported by \citet{hubrig2019,hubrig2019b}, we are unable to confirm the reported velocity variations, nor the sudden spectral and magnetic changes. 



\citet{hubrig2019b} also appear to conclude that the reports of the sudden spectral and magnetic change and the large velocity variations are erroneous. However, they ascribe this to a serious problem that they claim affects the FORS data reduction pipelines that they employed. Although we are unable to reproduce their experiment due to the proprietary nature of their data, we consider it highly unlikely that the large RV shifts and significant spectral distortions that they illustrate can result from the pipeline if it is properly used. Moreover, we point out that if these phenomena were indeed a consequence of a problem with the pipeline, this would strongly affect the magnetic measurements derived from all of their spectra of \HD. Apart from the observation obtained on February 17, 2018, this is not observed. Given this, in addition to the lack of detail about the data reduction and investigation into the purported problems, it seems that the simplest explanation is that human error, rather than the data reduction pipeline, is at fault.  

The FORS2 spectrum from February 17, 2018, reported by \citet{hubrig2019} to correspond to that of an early B star, is effectively identical to all of the other spectra of \HD\ in our re-reduced data, i.e. that of a late O-type star. In addition, the longitudinal magnetic field measured from the re-reduced spectrum, equal to $-250 \pm 140$\,G, is fully compatible with the field measured from the other spectra of this star. Examining figures 4 and 6 of \citet{hubrig2019} (in the latter of which they illustrate the remarkable similarity of the affected spectrum with that of the known magnetic B2 star CPD\,-57$\degr$\,3509), it seems reasonable to speculate that the spectrum of another object (in particular a rapidly rotating magnetic B-type star) may well have been mistakenly substituted by those authors. 

The new high-resolution spectropolarimetric observations of \HD\ presented here paint a clear picture of a very slowly rotating, strongly magnetized object, the general behaviour of which is compatible with the known sample of hot, magnetic stars. In particular, our most recent ESPaDOnS observation reveals a change of polarity of the longitudinal magnetic field implying that the northern magnetic hemisphere has now become visible.

\section*{Acknowledgments}
Based on observations obtained at the Canada-France-Hawaii Telescope (CFHT) which is operated by the National Research Council of Canada, the Institut national des sciences de l'Univers of the Centre national de la recherche scientifique of France, and the University of Hawaii. MES acknowledges the financial support provided by the Annie Jump Cannon Fellowship, supported by the University of Delaware and endowed by the Mount Cuba Astronomical Observatory. GAW acknowledges support from the Natural Sciences and Engineering Research Council (NSERC) of Canada in the form of a Discovery Grant. ADU acknowledges support from NSERC. This work has made use of the VALD database, operated at Uppsala University, the Institute of Astronomy RAS in Moscow, and the University of Vienna.

\bibliographystyle{mnras}
\bibliography{article}
\bsp	
\label{lastpage}
\end{document}